\documentclass[conference]{IEEEtran}
\usepackage{amsmath}
\usepackage{amsthm}
\usepackage{multirow}
\usepackage{graphicx}
\usepackage{url}
\newtheorem*{definition}{Definition}
\begin{document}
\title{Where in the Internet is congestion?}

%\author{\IEEEauthorblockN{Michael Shell}
% \IEEEauthorblockA{School of Electrical and\\Computer Engineering\\
% Georgia Institute of Technology\\
% Atlanta, Georgia 30332--0250\\
% Email: http://www.michaelshell.org/contact.html}
% \and
% \IEEEauthorblockN{Homer Simpson}
% \IEEEauthorblockA{Twentieth Century Fox\\
% Email: homer@thesimpsons.com}
% \and
% \IEEEauthorblockN{James Kirk\\ and Montgomery Scott}
% \IEEEauthorblockA{Starfleet Academy\\
% San Francisco, California 96678-2391\\
% Telephone: (800) 555--1212\\


\author{\IEEEauthorblockN{Daniel Genin
\IEEEauthorrefmark{1},
Jolene Splett
\IEEEauthorrefmark{2}
}
\IEEEauthorblockA{
\IEEEauthorrefmark{2}
Stastical Engineering Division\\
National Institute of Standards and Technology
%Atlanta, Georgia 30332--0250\\ Email: see http://www.michaelshell.org/contact.html
}
\IEEEauthorblockA{\IEEEauthorrefmark{1}
Johns Hopkins Applied Physics Laboratory
%National Institute of Standards and Technology
%Telephone: (800) 555--1212, Fax: (888) 555--1212}
%\IEEEauthorblockA{\IEEEauthorrefmark{4}Tyrell Inc., 123 Replicant Street, Los Angeles, California 90210--4321}
}
}


\maketitle

\begin{abstract}
%\boldmath
Understanding the distribution of congestion in the Internet is a long-standing problem. Using data from the SamKnows US broadband access network measurement study, commissioned by the FCC, we explore patterns of congestion distribution in DSL and cable Internet service provider (ISP) networks. Using correlation-based analysis we estimate prevalence of congestion in the periphery versus the core of ISP networks. We show that there are significant differences in congestion levels and its distribution between DSL and cable ISP networks and identify bottleneck sections in each type of network. 
\end{abstract}


% \fi
\IEEEpeerreviewmaketitle

%%%%%%%%%%%%%%%%%%%%%%%%%%%%%%%%%%%%%%%%%%%%%%%%%%%%%%%%%%%%%%%%%%%%%%%%%%%%%%%%%%%%%%%%%%%%%%%%%%%
\section{Introduction}

% Distribution of Internet congestion is an important and difficult problem
Internet congestion has been a topic of active research for as long as the Internet has existed\cite{BauClaLeh}. Congestion in computer networks more generally has been studied even longer and is itself predated by research on congestion in telephone networks and car traffic. There are good reasons for being interested in Internet congestion since it largely determines everyday Internet user experience from the time it takes to load a webpage, to the visual quality of streaming media and responsiveness of the on-line gaming experience. One important  question concerns the distribution of congestion in the Internet\cite{CroKri}. Is it mainly concentrated in the so-called ``core" of the network or at its edge or somewhere in between? Is it evenly distributed or concentrated at the edge in some networks and in the core in others? Perhaps congestion can occur in any of those network segments at different times. The answers to these questions have so far proved elusive\cite{BauClaLeh}. The companies that own various parts of the network guard their traffic, capacity, and topology data to maintain their competitive advantage, making it virtually impossible for the research community to obtain any kind of insight into the congestion distribution. Yet, even the Internet service providers (ISPs) would benefit from such understanding as it would allow them to target infrastructure improvements at the key points in the network where return on investment, in terms of enhanced user experience, would be greatest.

%Scarcity of data
To date, the scarcity of data has been one of the main obstacles to understanding congestion in the Internet. Simulations can shed some light on the range of possible network congestion regimes, but in the absence of real data it is hard to say what is and is not likely to be observed in practice\cite{FloKoh}. In other words, simulations circumscribe the state space but provide no probability distribution describing the likelihood of various states. This is not surprising because in order to assign such probabilities one would have to know something about the distribution of demand and about routing, which is precisely the information we lack.

The data that would permit research to move forward is also hard to collect due to lack of measurement infrastructure. Understanding the distribution of congestion requires, at a minimum, end-to-end measurement of a range of characteristics on a large and representative sample of Internet paths\cite{CroKri}. On the other hand, the Internet has hundreds of millions of end-nodes\cite{Aka}, the vast majority of which are individual user devices that can only be used for measurement purposes with the consent of the user. One approach to tackling this problem
%,implemented by Ookla Speedtest\cite{Ook} and M-Lab Netalyzr\cite{KreWeaNecPax}, 
is to set up a limited number of measurement servers and to allow users to test their connections against these servers when they choose to (see, for example, \cite{Ook},\cite{KreWeaNecPax}). Unfortunately, this approach has several drawbacks if they are to be used for statistical analysis. 
\begin{enumerate}
\item The samples are not collected to be representative in terms of network distribution or geography. This can potentially be dealt with post factum by paring down the data sets to create a representative sample. 
\item Dynamic allocation of Internet protocol (IP) addresses by ISPs makes it impossible to ensure that measurements corresponding to the same IP address (which is the only unique connection identifier) truly correspond to the same physical connection.
\item Since each measurement must be initiated by the user, they are performed at irregular intervals and may also be biased toward times when the network performance is particularly poor.
\item Because measurement code runs on the user's computing device, the measurement is subject to interference from the device's operating system environment as well as other devices in the user's home network.
\end{enumerate}
Thus, while such data sets provide a glimpse of  network performance from the end-user perspective, their value for robust inferences about Internet congestion is limited.

In 2010 the Federal Communications Commission (FCC)
%, as part of implementation of the National Broadband Plan\cite{Fcc}, 
contracted with a private company, SamKnows, to perform the first large scale measurement study of America's broadband access networks. This study, still on-going, is the largest such effort to date, with over 10,000 measurement units deployed on ISP customers' premises and 16 ISPs participating in the study. Moreover, the measurement approach taken by SamKnows suffers none of the issues mentioned above. Section \ref{data_and_definitions} describes the details of the data collection methodology relevant to the present manuscript. A full description of the data collection and sampling methodology can be found in ``Measuring Broadband America, Technical Appendix" by SamKnows \cite{Sam}.

%our contribution
We analyze the publicly available SamKnows data to identify the location of congested network resources in the US broadband access network\footnote{Any mention of commercial products is for information only; it does not imply recommendation or endorsement by NIST.}. Our approach is two-fold. First, we identify connections that experience substantial performance degradation. Then we use correlation-based analysis for the detection of shared congested resources to determine where the congestion occurs relative to the customer's side of the connection. 

The rest of the paper is organized as follows. Relevant details of the SamKnows data collection methodology and the definitions necessary for data analysis are laid out in Section \ref{data_and_definitions}. The data processing and analysis methodologies are described in Section \ref{methodology}. Results of the analysis are described in Section \ref{results}, followed by a discussion of open issues in Section \ref{discussion}. Section \ref{conclusion} summarizes our findings and outlines directions for future research.

%%%%%%%%%%%%%%%%%%%%%%%%%%%%%%%%%%%%%%%%%%%%%%%%%%%%%%%%%%%%%%%%%%%%%%%%%%%%%%%%%%%%%%%%%%%%%%%%%%%
\section{Data and definitions}
\label{data_and_definitions}
Our study is based on the publicly released data collected by SamKnows during the period March through June 2011\cite{Sam1}. The collected dataset comprises hourly measurements of 13 network characteristics measured from 13,404 measurement units deployed across 16 ISPs. 

There are two aspects of the SamKnows study that set it apart from previous residential broadband measurement projects: first, the meticulous positioning of measurement units to ensure coverage across geographical regions and ISP networks and, second, the deployment of advanced measurement hardware that insulates measurements from most\footnote{There is some evidence that characteristics of cable/DSL modems, the one device between the measurement unit and the ISP's line, may affect measurements\cite{SunDonFeaTeiCraPes}.} of the effects of a consumer's networking and computing equipment\cite{Sam}.

The SamKnows measurement infrastructure consists of measurement units and measurement servers. Measurement units, also referred to as Whiteboxes by SamKnows, are Netgear wireless routers with custom built firmware that incorporate code for running SamKnows measurement tests. Measurement servers are hosts used as measurement reference points. The primary measurement servers are located at nine geographically distributed locations and hosted by M-Lab\cite{Sam}. Additionally, some of the participating ISPs host secondary measurement servers inside their own networks for verification purposes.

The measurement units were configured to run a battery of 13 tests at one or two hour intervals (depending on the test) starting with some random time offset, to balance the load on measurement servers. The tests included multi-threaded transmission control protocol (TCP) download and upload throughput benchmarks, download timings for front webpages of ten popular websites, round-trip latency measurements and a range of other tests\cite{Sam}. In the present study we focus our attention on the first two tests: the multi-threaded TCP download speed benchmark and website download speed test, because these measurements include a diversity of network paths that allows us to make deductions about the distribution of congestion in broadband access networks.

The multi-threaded TCP download throughput benchmark consisted of an extended TCP data transfer session. Three concurrent TCP connections were used to ensure saturation of the available bandwidth. The transfer of the main payload was preceded by a warm-up period which consisted of 
\begin{quote}``repeatedly transferring small chunks (256 kilobytes, or kB) of the target payload before the real testing began. This `warm-up' period was said to have been completed when three consecutive chunks were transferred at within 10\% of the speed of one another. All three connections were required to have completed the warm up period before the timed testing began. The `warm-up' period was excluded from the measurement results."\cite{Sam} 
\end{quote}
\noindent The download speed test continued for 30 seconds with byte totals transferred recorded at 5 second intervals. The download speed test was performed every two hours.

The website download speed test measured the time necessary to download the front webpage of each of the first ten \emph{Alexa US Top 500 Global Sites}: 
\begin{description}
\item http://www.cnn.com, 
\item http://www.youtube.com, 
\item http://www.msn.com, 
\item http://www.amazon.com, 
\item http://www.yahoo.com, 
\item http://www.ebay.com, 
\item http://www.wikipedia.org, 
\item http://www.facebook.com, 
\item http://www.google.com, 
\item http://www.netflix.com. 
\end{description}
``The primary measure for this test was the total time taken to download the HTML front page for each website and all associated images, JavaScript, and stylesheet resources."\cite{Sam}.

We now lay out the definitions necessary to rigorously frame the subsequent discussion. The first definition is related to the download performance of a given connection. 
\begin{definition}
Let $X=\{X_i\}_{i=1}^N$ be a sequence of the TCP download throughput measurements for a given connection. We will say that the connection experiences \textbf{\emph{ $\bf (q, t)$-recurrent congestion}} if
\begin{equation*}
P\left[\frac{X}{X_{max}}<q\right]>t
\end{equation*}
where $P[]$ stands for the fraction of samples satisfying the given condition (and is meant to be suggestive of probability), $X_{max}$ is the ISP assigned download speed tier\footnote{This is the upper limit of download speed, usually stated as ``up to" speed, specified in the customer-ISP contract.} of the connection, $q\in[0,1]$ is the fraction of the speed tier attained and $t\in[0,1]$ is the fraction of the time the measurement is below $qX_{max}$.
\end{definition}
This definition covers the full range of performance levels from nearly perfect $(.99,0)$ to complete failure $(0,100)$. We refer to this performance measure as recurrent congestion because it quantifies the performance level as well as its duration. In particular, we will be interested in connections whose low performance occurs a significant fraction of the time, hence recurrent congestion.

\begin{figure*}[ht]
\begin{center}
\includegraphics[height=5cm]{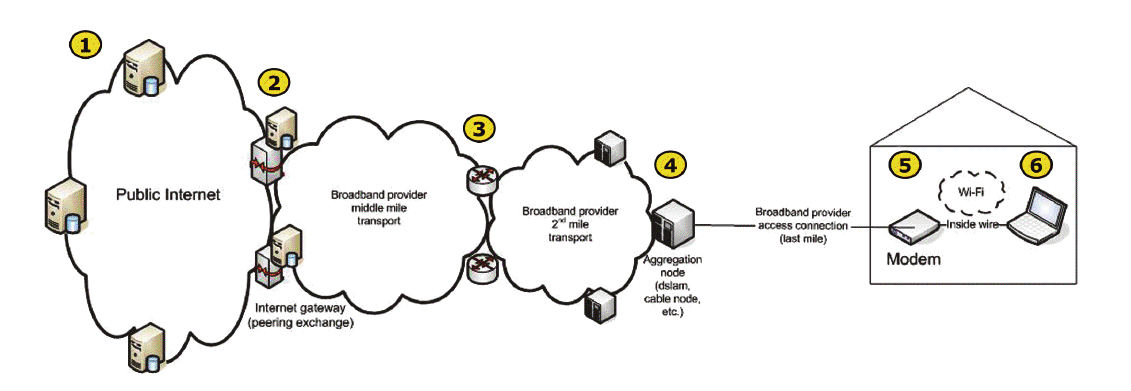}
\end{center}
\caption{Boundary demarcations for residential broadband connections}
\label{fig_internet_schematic}
\end{figure*}

Next, we introduce the notion of initial segment.
\begin{definition}
\label{def_initial_segment}
Let $P_t(d)$ be the path taken by packets leaving a given measurement unit at time $t$ for destination $d$, expressed as a sequence of IP level router interfaces (such as produced by traceroute). Let $P(d)=\cup_{t\in[T_0,T_1]} P_t(d)$, where $T_0$ and $T_1$ specify the time interval of interest, then \textbf{\emph{initial segment}} of the connection corresponding to the given measurement unit is
\begin{equation}
I=\bigcap_{d\in\mathcal{D}}P(d),
\end{equation}
where $\mathcal{D}$ is the collection of destinations external to the ISP of the connection.
\end{definition}
The initial segment is clearly well defined since all the sets involved in the definition are finite, but may theoretically turn out to be the empty set. At the same time, the initial segment is not necessarily a collection of consecutive links but, depending on the routing architecture of the ISP, may turn out to be a subnetwork or even a collection of disjoint segments. As defined, the initial segment loosely corresponds (somewhat ironically) to the union of the ``last"  and ``second mile" connectivity in the standard user centered network segmentation scheme (Figure \ref{fig_internet_schematic}\cite{Fcc}) and certainly does not extend beyond the ``peering exchange" point.

The union over time, in the definition of the initial segment, is necessary to account for routing changes on the time scale of the time series considered. Clearly, repeated computations of the initial segment over a time period long compared with $[T_0,T_1]$ may give different answers. For instance, the initial segment computed over the span of a month may change significantly on the time scale of years due to changes in the physical network infrastructure and it may change on the time scales of months and weeks due to changes in routing policies. This said, we believe that the initial segment is sufficiently stable over a period of a month to be considered a time invariant construct.

In practice, the initial segment of a given end host can be relatively easily computed using \emph{traceroute} and a representative sample of destinations. Unfortunately, \emph{traceroute} mapping was not among the tests run by SamKnows and so we cannot explicitly describe the initial paths of the connections in the study.

The remaining ingredient of our study is the notion of a tight initial segment. 
\begin{definition}
A network resource, e.g., a link or a router, is said to be \textbf{\emph{tight}} (or \textbf{\emph{narrow}}) at some fixed moment in time for a given traversing path if its available bandwidth determines the bandwidth of the whole path\cite{JaiDov}.
\end{definition}
It is clear that every Internet path has a unique tight segment. Note that the tight resource is path dependent, i.e., a resource that is tight for one path may not be tight for another.
\begin{definition}
An \emph{initial segment is tight} at a fixed time $t$ if the initial segment contains a network resource that is tight at time $t$ for at least one path terminating beyond the initial segment.
\end{definition}
The initial segment includes all network devices between consecutive IP router interfaces, i.e. all network devices and links between the users side of the connection and the terminal node(s) of the initial segment. A tight initial segment need not be tight for all paths through the Internet. It is quite likely that a long inter-continental path will be constrained somewhere beyond its initial segment. 
%Furthermore, although the definition is not parametrized, it is clear that some tight initial segments will impose a higher performance penalty than others. So a {\it slightly} tight initial segment may constrain the throughput of only a few paths, while a {\it very} tight initial segment may be limiting for virtually all paths.

%%%%%%%%%%%%%%%%%%%%%%%%%%%%%%%%%%%%%%%%%%%%%%%%%%%%%%%%%%%%%%%%%%%%%%%%%%%%%%%%%%%%%%%%%%%%%%%%%%%
\section{Methodology}
\label{methodology}

We now discuss how recurrent congestion and tightness of initial segments can be estimated based on the SamKnows measurement data.

Collecting data on a large scale is bound to meet with some technical difficulties. Whiteboxes located in customers' homes or businesses might lose connectivity for any number of reasons, from an ISP network outage to a power outage or even user interference. So it is not surprising that the data set contained a significant number of missing measurement runs.  We required that the data for individual connections have a sufficiently large number of matched multi-threaded TCP and website download speed measurements. Thus any connection that had fewer than half of the roughly 12*30=360 total possible monthly matching pairs of measurements was dropped from consideration. This reduced the number of connections available for analysis from about 13,000 to 3,000.

Recurrent congestion is fairly easy to compute since SamKnows data includes multi-threaded TCP download throughput benchmarks. The only difficulty lies in determining a connection's speed tier. SamKnows provides information on the speed tiers of connections but only as they were specified at the time the study was initiated. The dataset, on the other hand, spans four months during which some of the participants in the study modified their broadband subscriptions and changed providers. 

Based on the available data, there is no way to detect a change of ISP for a given connection, though one might expect a change in ISP to be reflected in the change of the speed tier, especially since speed tier structures of different ISPs do not always match. Changes in the connection's speed tier can be detected since they typically lead to an abrupt change in the maximum attainable download throughput. To detect the change in maximum achievable download throughput of a given connection, we applied the following algorithm.
\begin{enumerate}
\item For every day in a given month we selected the maximum sustained\footnote{To deal with dynamic bandwidth technologies such as PowerBoost only the last 5 seconds of the download speed test were used\cite{Sam}.} download throughput achieved on that day.
\item If the largest and smallest of the daily maxima differed by more than 50\% of the mean the connection was declared to have changed speed tiers.
\item If the connection's speed tier did not change then its speed tier was set to the mean of the daily maximum sustained download speeds during the given month.
\end{enumerate}
Connections detected to have changed their speed tiers were dropped from further consideration. This reduced the number of connections available for analysis by roughly 5-7\%.

%Next, a discrete download speed distribution was computed for each connection by binning sustained download speeds by rounding down to the nearest tenth , the fraction of the speed tier attained during each test run.   Such discrete distributions make it easy to check whether a connection was recurrently congested for $q$ parameter restricted to tenths fractions.  [j - I'm a little confused by this paragraph.]

The complete two-parameter definition of recurrent congestion is particularly useful when exploring the functional dependence of a connection's download throughput performance on other factors, but it is not always convenient to specify $(q,t)$ parameters at every reference to recurrent congestion when discussing congestion as a network state. Since we will not be exploring the functional form of dependence, it will be useful to settle on some canonical threshold at which the connection can be said to experience recurrent congestion. We will set this threshold at what might be considered the ``C" grade level of $(.8,.2)$-recurrent congestion, i.e. less than 80\% of the speed tier measured more than 20\% of the time. While 20\% of a month's span may not sound like much, when considered as a fraction of peak usage hours\footnote{Peak usage hours are weekday hours between 7pm and 11pm\cite{Sam}.} (during which the download speed is most likely to be degraded), it works out to be almost 90\%.

Detecting tight initial segments is considerably more complicated because tightness of an initial segment is not something directly measurable. Rather, an indirect method, relying on correlation between download speed measurements for paths sharing the same initial segment  will be used.

\begin{figure}[ht]
\begin{center}
\includegraphics[width=6cm]{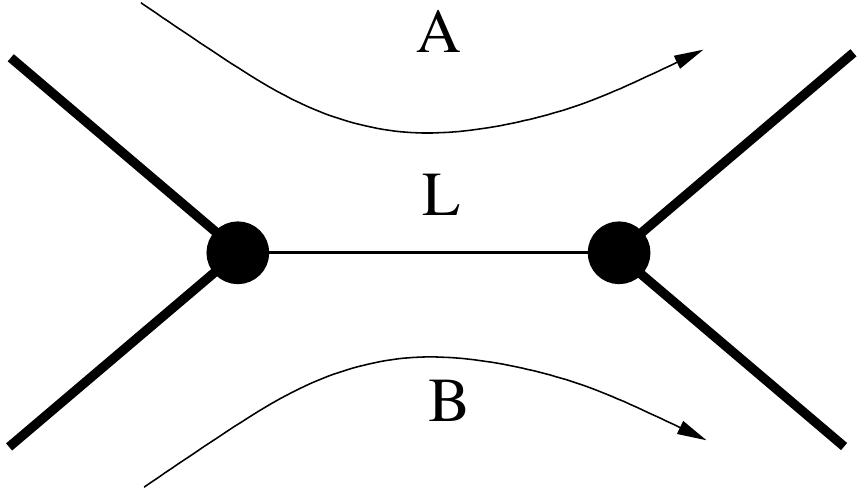}
\end{center}
\caption{A tight link $L$ shared by two flows $A$ and $B$}
\label{fig_tight_link}
\end{figure}

The idea of using correlation to discover shared congested resources is not new\cite{RubKurTow},\cite{HarBesBye},\cite{PadQiuWan},\cite{Duf}. The basic principle underlying all these approaches is that flows constrained by the same network resource are likely to exhibit temporal correlation in their characteristics. For example, suppose that two TCP flows, $A$ and $B$, share a tight link $L$ (Figure \ref{fig_tight_link}), i.e., a link whose available bandwidth determines the throughput of the two flows. As the available bandwidth of link $L$ changes over time, so too will the throughput of flows $A$ and $B$. Since under TCP throughput is proportional to the available bandwidth, the measured throughputs of the two flows will be synchronized (at least as long as $L$ remains tight) and will exhibit temporal correlation. 

We detected connections with tight initial segments by adapting this correlation-based approach to the SamKnows measurement data. Since all paths originating from a given connection have some section, including the initial segment, in common, correlation in download speeds on any pair of paths gives some information about potential tightness of the initial segment. The more pairs of paths exhibit a high correlation, the more likely it is that the initial segment is tight. Indeed, while a few pairs may by chance exhibit high correlations independently due to tightness on links beyond the initial segment, this becomes increasingly unlikely as the number of pairs considered increases.

The SamKnows dataset contains measurements for download speed benchmark as well as website download speeds to which the outlined method can be applied. What distinguishes the present setting from prior applications of this idea is that the measurements are not strictly simultaneous and that the interval between consecutive measurements is hours rather than seconds or milliseconds. Since pairs of paths could not be measured at exactly the same time, we assumed that pairs of paths measured relatively close together in time were measured simultaneously. This assumption is reasonable given that the separation between measurement runs is roughly two hours. The long interval between measurement runs drastically limits the resolution of the correlation method. Correlating such sparse measurements can only reveal tight network resources that remain congested for hours. Sporadic tightness on the order of minutes or tens of minutes will be invisible to our detection method.

Details of the SamKnows measurement methodology also had to be taken into account. Caution had to be taken to avoid spurious correlation, i.e. one not actually resulting from a tight resource on the shared portion of the paths. The website download speed measurements may exhibit correlation due to the diurnal user activity pattern. Since websites tend to experience heavier loads during peak usage hours, download speeds across many websites may be affected simultaneously, causing correlation when there is no shared tight resource in the path. To avoid this effect, we considered only pairwise correlations between the download speed benchmark and website download speed measurements, but not between website download speed measurements themselves. Spurious correlation between a download speed benchmark, performed against a measurement server, and a website download speed measurement is highly improbable. This gives ten pairs of paths, where each pair consists of a path to one of the websites and an M-Lab measurement server.

In order for the correlation to be useful, the paired series should have a sufficiently large number of measurements. There is some leeway in defining ``sufficiently large" and we settled on 180 measurements, which is half of the 360 total possible. Limiting analysis to connections with 180 or more paired measurements reduced the number of connections available for analysis from about 13,000 to roughly 3,000. Setting the limit much higher would have reduced the number of connections even further.

Finally, for the connections that passed muster, correlations for each of the ten pairs of paths were computed for each month, yielding roughly 3,000$\times$10$\times$4 correlations. To decide when a given connection has a tight initial segment we chose a high correlation threshold of 0.6 and a high correlation count threshold of 5. Every connection that had 5 or more correlations greater than 0.6 was declared to have a tight initial segment. These cut-offs are again a somewhat arbitrary mark on a continuum scale, however, they are sufficient to provide an idea of the relationship between initial segment tightness and recurrent congestion, which is what we are really after. We expect that the number of connections with tight initial segments varies ``continuously" with the high correlation and high correlation count threshold, by which we mean that a small change in either or both quantities would lead to a small change in the number of connections with a tight initial segment.

%%%%%%%%%%%%%%%%%%%%%%%%%%%%%%%%%%%%%%%%%%%%%%%%%%%%%%%%%%%%%%%%%%%%%%%%%%%%%%%%%%%%%%%%%%%%%%%%%%%
\section{Results}
\label{results}

There are many interesting observations that can be made on the basis of the outlined analysis, but since our main concern in the present paper is the distribution of congestion we will limit our exposition to observations most salient to this topic.

We consider only DSL and cable connections since the data set did not include a sufficient number of measurements for analysis of wireless and fiber-optic technologies. Our study indicates that cable and DSL technologies have substantially different performance in the sense of delivering bandwidth at the assigned speed tier as well as in distribution of the congestion.

In this data set, DSL broadband provided connections on average delivering download speeds above 80\% of the assigned speed tier more than 80\% of the time. By contrast, a significant fraction of cable connections received less than 80\% of their assigned speed tier more than 20\% of the time. One must keep in mind that cable connections typically have higher download speed tiers than DSL connections. In the SamKnows data the average download speed tier for DSL connections was 5.4 Mbps vs. 13.5 Mbps for cable connections.

\begin{figure}[h]
\begin{center}
\includegraphics[width=8.5cm]{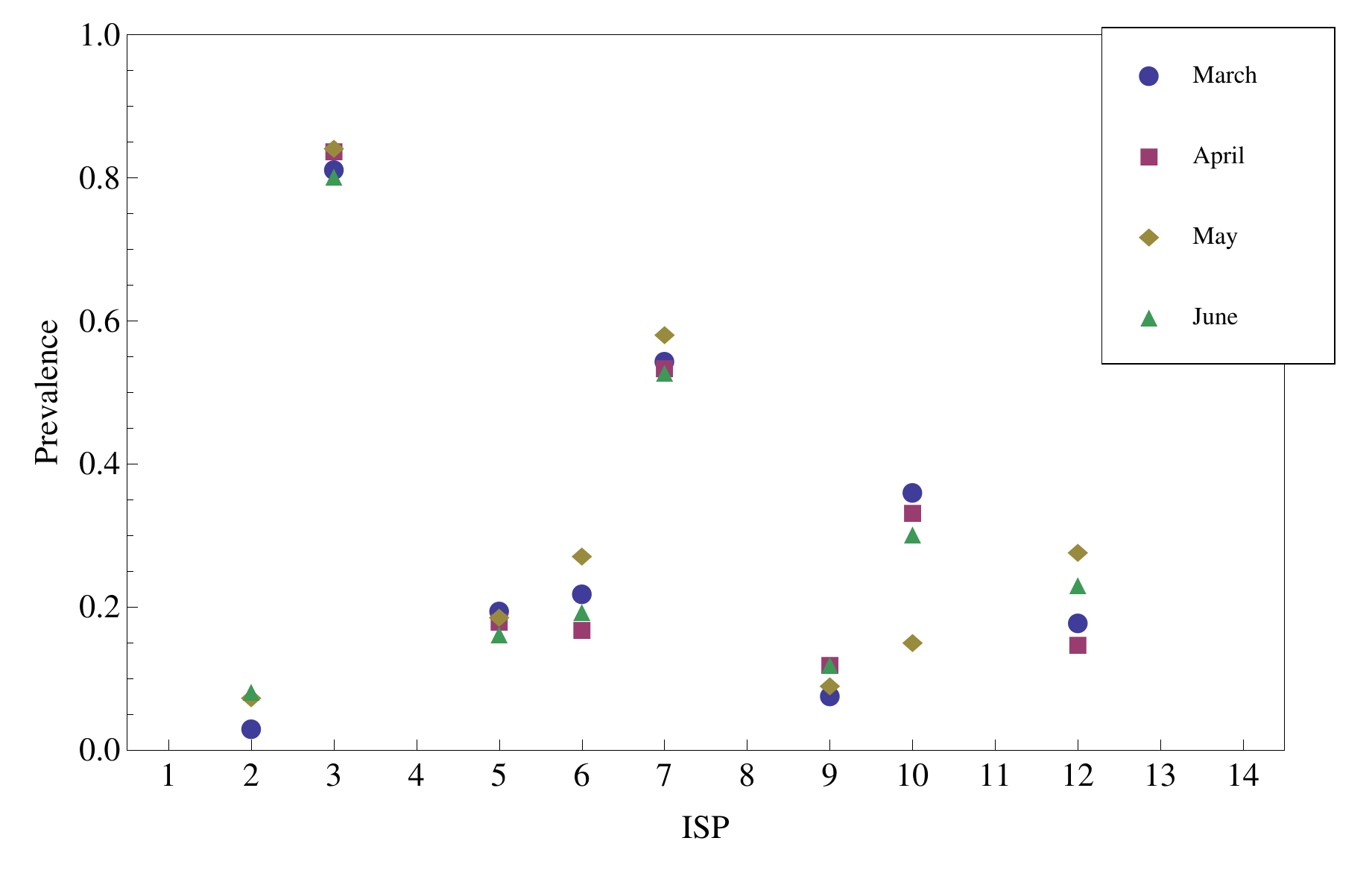}
\end{center}
\caption{Recurrent congestion prevalence for cable connections by ISP}
\label{fig_cable_rc_prevalence_by_isp}
\end{figure}

\begin{figure}[h]
\begin{center}
\includegraphics[width=8.5cm]{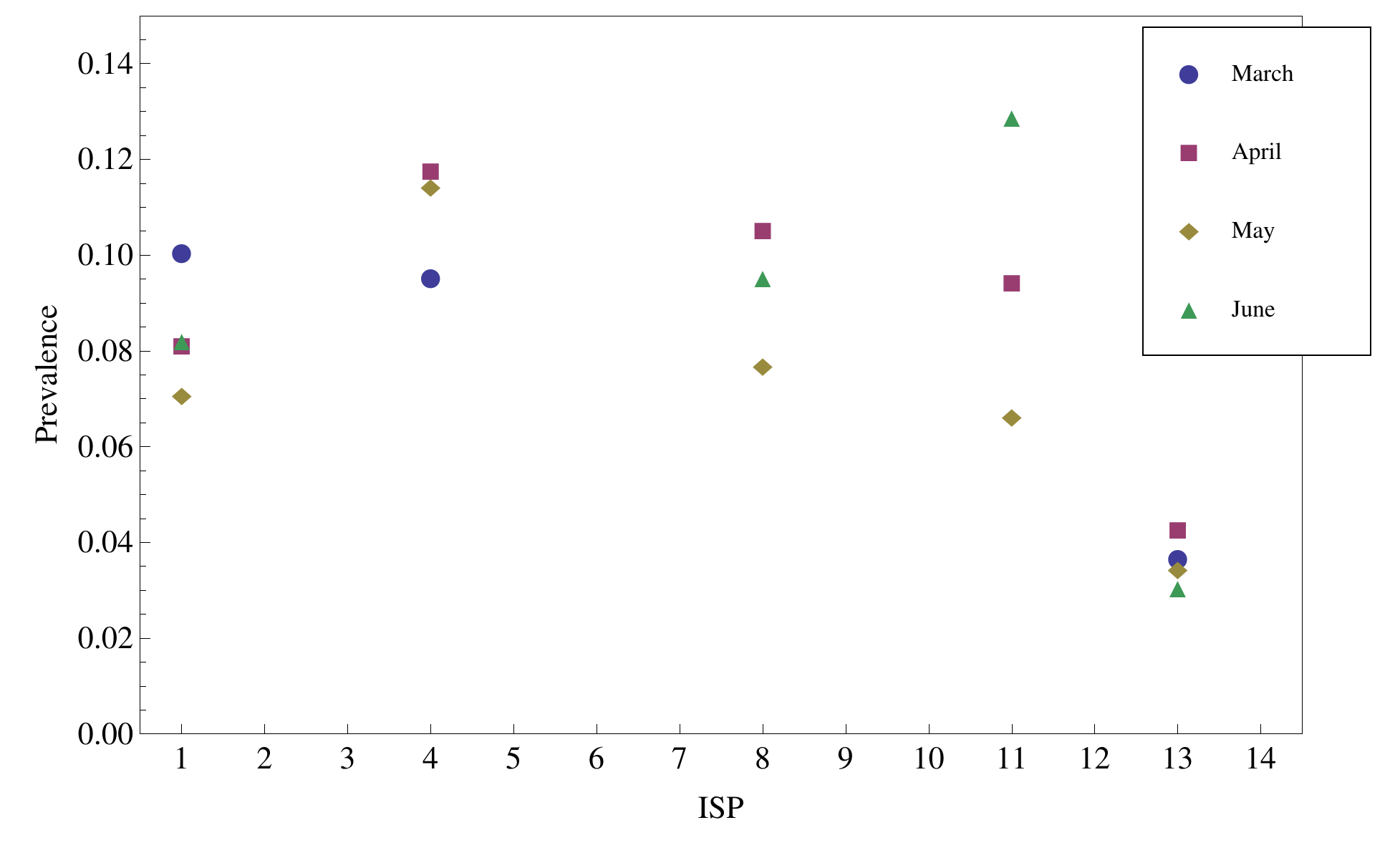}
\end{center}
\caption{Recurrent congestion prevalence for DSL connections by ISP}
\label{fig_dsl_rc_prevalence_by_isp}
\end{figure}

Tables \ref{table_cable} and \ref{table_dsl} summarize recurrent congestion (RC) and tight initial segment (TIS) data for cable and DSL samples respectively. The difference in consistency of service is reflected in the number of connections with recurrent congestion, a relatively low 9--12\% for DSL in comparison to 27--32\% for cable connections. The distributions of recurrently congested connections are very different among the ISPs (Figures \ref{fig_cable_rc_prevalence_by_isp} and \ref{fig_dsl_rc_prevalence_by_isp}). For DSL ISPs, concentrations of recurrently congested connections are roughly similar, for cable connections there are several ISPs that have disproportionately high concentrations of recurrently congested connections. As will be discussed further, this is closely connected with the prevalence of tight initial segments in the poorly performing ISP networks.

\begin{table}[h]
\begin{center}
\begin{tabular}{|l|r|r|r|r|r|r|}
\hline
Month 					& Total & FC	 & PTIS	& FC$\cap$PTIS &FC$\cap$PTIS/PTIS & FC$\cap$PTIS/FC\\
\hline
\multirow{2}{*}{March} 	& 1559 	& 416  	 & 55 	& 55 	& 100\%	& 13\% \\
						&		& 27\%	 & 4\% 	& 		&		&\\
\hline
\multirow{2}{*}{April} 	& 1864 	& 454 	 & 57 	& 52 	& 91\%	& 11\% \\
						& 		& 24\% 	 & 3\% &		&		&\\
\hline
\multirow{2}{*}{May} 	& 1818  & 582  	 & 49 	& 45 	& 92\% 	& 8\% \\
						& 		& 32\% 	 & 3\% 	&		&		&\\
\hline
\multirow{2}{*}{June}	& 1903 	& 519  	 & 51  	& 50 	& 98\% 	& 10\%\\
						& 		& 27\%	 & 3\%	&		&		&\\			
\hline
\end{tabular}
\caption{Cable data}
\label{table_cable}
\end{center}
\end{table}

\begin{table}[h]
\begin{center}
\begin{tabular}{|l|r|r|r|r|r|r|}
\hline
Month 					& Total & FC	 & PTIS	& FC$\cap$PTIS &FC$\cap$PTIS/PTIS & FC$\cap$PTIS/FC\\
\hline
\multirow{2}{*}{March}	& 860 	& 99  	& 55  	& 37 	& 67\% 	& 37\% \\
						&		& 12\% 	& 6\%	&		&		&\\
\hline
\multirow{2}{*}{April}  & 1031 	& 103  	& 75 	& 50 	& 67\% 	& 49\% \\
						&		& 10\%	& 7\%	&		&		&\\
\hline
\multirow{2}{*}{May} 	& 1059 	& 91 	& 56 	& 38 	& 68\% 	& 41\% \\
						&		& 9\%	& 5\%	&		&		&\\
\hline
\multirow{2}{*}{June}	& 1012 	& 110 	& 68 	& 44 	& 65\% 	& 40\% \\
						&		& 11\%	& 7\%	&		&		&\\
\hline
\end{tabular}
\end{center}
\caption{DSL data}
\label{table_dsl}
\end{table}

The percentages of connections with tight initial segments for DSL --- 5--7\% --- are roughly double those of cable connections --- 3--4\%. There are also major differences when recurrent congestion and tightness of initial segments are considered together.

In the case of DSL 37--50\% of the connections identified as having a tight initial segment also experienced recurrent congestion, whereas for cable connections the same number was 91--100\%. That is, a tight initial segment virtually always coincides with recurrent congestion for cable connections but more than half of DSL connections manage to deliver performance close to speed tier in spite of a tight initial segment. This difference also underscores the point that a tight initial segment need not lead to severe performance degradation.

More interesting is the percentage of recurrently congested connections that also have tight initial segments. While a tight initial segment need not coincide with recurrent congestion, when the two do coincide we can conclude that recurrent congestion is most likely due to congestion on the initial segment. If observed recurrent congestion was not due to congestion on the initial segment the correlation between different download speed measurements could only be explained by a chance coincidence, making it extremely unlikely\cite{Duf}.

For DSL connections 65--67\% of recurrently congested connections also had tight initial segments whereas for cable connections this percentage was only 8--13\%. Taken at face value these numbers seem to suggest that a significant amount of congestion, especially for cable connections, occurs deeper in the network, perhaps, in the ``middle mile" (Figure \ref{fig_internet_schematic}) or even farther, where the ISP connects to the ``public Internet". This is somewhat contrary to the popular belief that the edge is more congested than the core\cite{PadQiuWan}. Here, however, we must be careful not to exceed the limitations of our congestion localization approach. It is possible that because our method for detecting tight initial segments underestimates their prevalence, they could still be the dominant cause of recurrent congestion while remaining undetected.

To the extent that paths to a few popular websites may turn out to have a congestion distribution that is substantially different from the majority of the paths originating from the same point, our estimates may give the wrong picture about the presence of congestion at the periphery of the ISP networks. This may happen, for example, because correlation due to tightness of initial segment is most likely to arise at times of peak usage, which are also the times when websites are under heaviest load, possibly causing reduction in download speeds due to limited bandwidth near the site itself. 

We have no way to verify directly the degree to which our analysis is impacted by such effects but we can look for signatures that these effects may leave in the results. Significant insight can be gained by considering variation in recurrent congestion and tight initial segment prevalence across ISPs. Since ISP networks are largely autonomous, their internal patterns of congestion can be assumed to be independent. On the other hand, being interconnected mainly through the ``public Internet", ISPs are affected in roughly equal measure by congestion in the core of the Internet.

Consider the partitioning of the Internet from the point of view of a particular connection as in Figure \ref{fig_internet_schematic}. Since the M-Lab servers used for download throughput measurements were located in the ``public Internet" (relative to the ISPs that participated in the study) congestion affecting the download throughput could occur at or between any of the Points 1 through 4 in Figure \ref{fig_internet_schematic}.

Let us consider the possible scenarios under which tight initial segments are the primary limiters of download speed but are undetected because of atypical congestion on the paths to the test websites.

Suppose, first, that most connections were really constrained on the initial segment (points 3 and 4 in Figure \ref{fig_internet_schematic}) but that these tight initial segments went undetected because the website download measurements were constrained near the test website end of the path (point 1 in Figure \ref{fig_internet_schematic}). Since the bottlenecks at such locations in the Internet are likely to affect connections in different ISP networks equally, one would expect that the number of tight initial segments detected in an ISP's network would be roughly proportional to the number of the ISP's connections in the sample. 

\begin{figure}[h]
\begin{center}
\includegraphics[width=8.5cm]{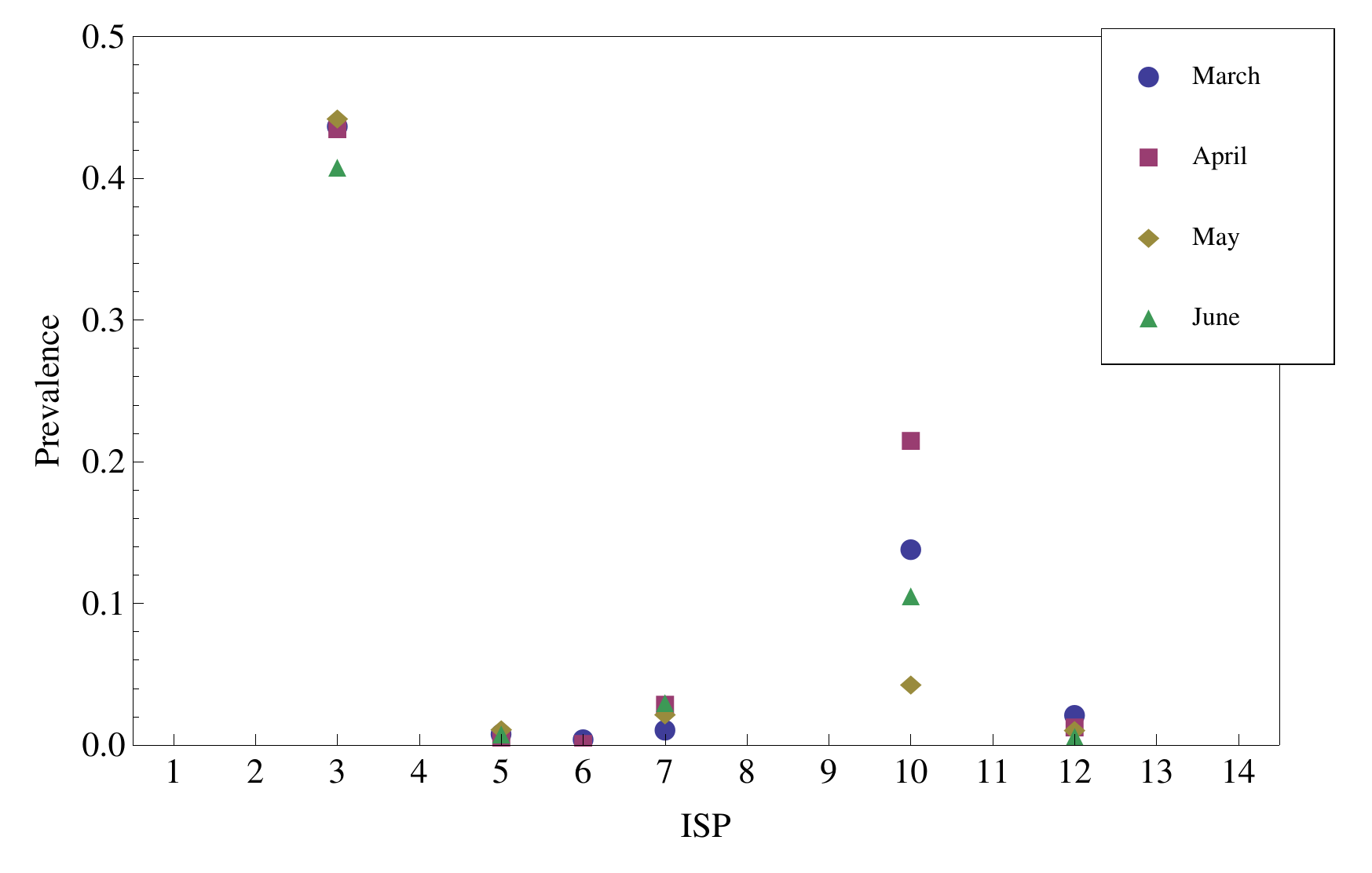}
\end{center}
\caption{Tight initial segment prevalence for cable connections by ISP \hspace{7em} In this and subsequent figures different colors correspond to different months. Blue for March, cyan for April, yellow for May and green for June.}
\label{fig_cable_tis_prevalence_by_isp}
\end{figure}

\begin{figure}[h]
\begin{center}
\includegraphics[width=8.5cm]{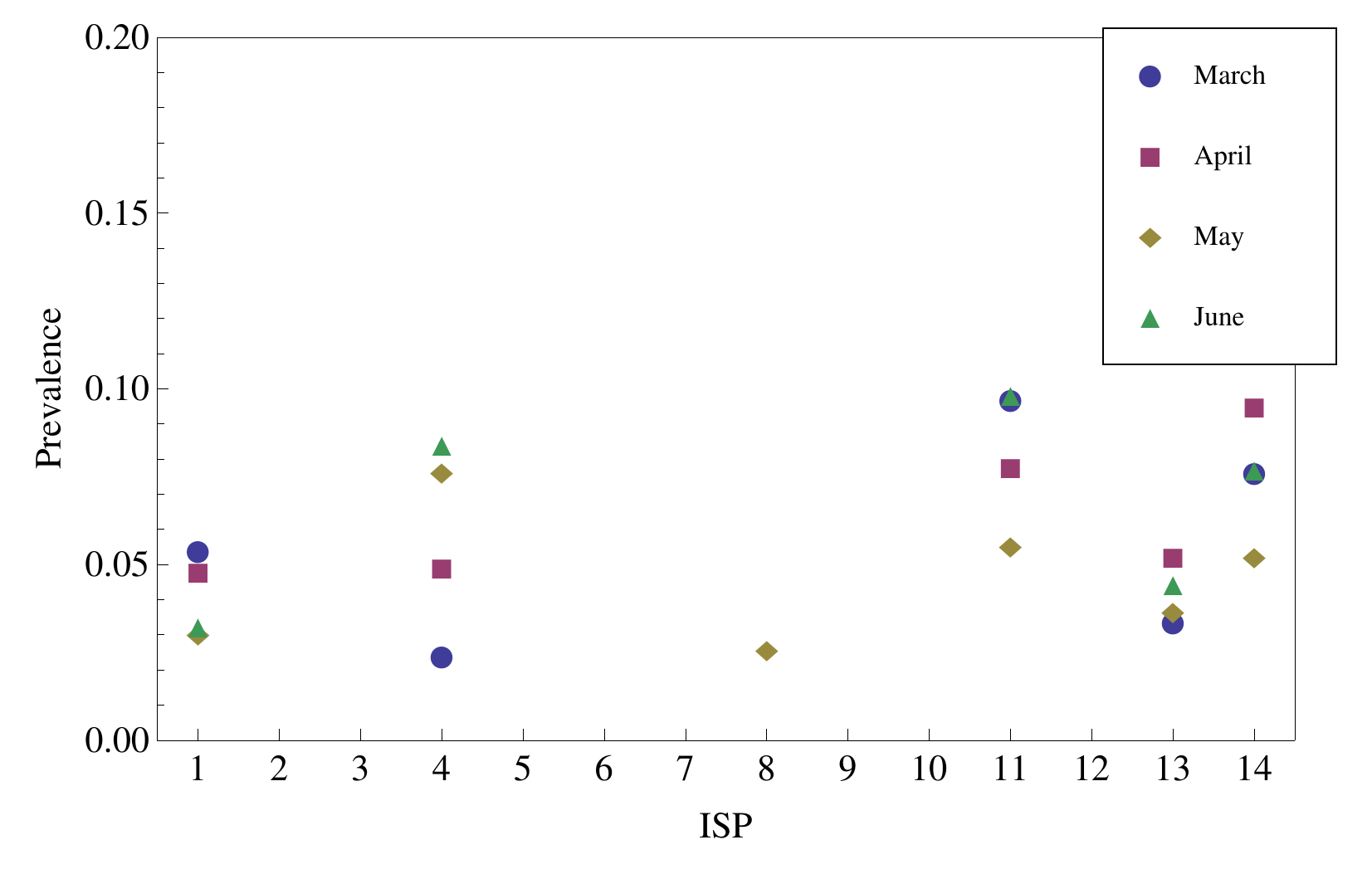}
\end{center}
\caption{Tight initial segment prevalence for DSL connections by ISP}
\label{fig_dsl_tis_prevalence_by_isp}
\end{figure}

Figures \ref{fig_cable_tis_prevalence_by_isp} and \ref{fig_dsl_tis_prevalence_by_isp} show the relative frequency of tight initial segments detected across ISPs for cable and DSL connections respectively. The plot for DSL connections appears to be consistent with the proposed hypothesis that most connections are constrained on the initial segment but that this fails to be detected because the paths to the chosen test websites encounter tight resources near the test websites.
%which is to some extent supported by the relatively high (at least as compared to cable connections) percentage of connections with tight initial segments experiencing recurrent congestion.

On the other hand, ISPs 3 and 10 (cable connections) have much higher concentrations of tight initial segments than the rest of the field, suggesting that either initial segment congestion is genuinely more prevalent in some ISP networks than others or that tight initial segments fail to be detected due to interference in the ``middle mile" (between points 2 and 3 in Figure \ref{fig_internet_schematic}) rather than near the test websites.

There is no reason to expect the rates of false negatives (i.e., undetected tight initial segments) for DSL and cable technologies should be significantly different. Although Figure \ref{fig_dsl_tis_prevalence_by_isp} is consistent with the hypothesis that congestion is predominantly concentrated on the initial segment, this possibility may be ruled out for cable,  and it also becomes less likely for DSL. In light of this, the more plausible scenario is that even as a large fraction of DSL connections is constrained by their initial segments, a significant fraction is also constrained elsewhere in the network.

% Initial segment congestion masked by congestion on website path on the segment Point 2 to Point 3.
Let us now consider the remaining possibility that connections are constrained primarily on their initial segments but that this fails to be detected because website measurements are hit by congestion in the ``middle mile" between Points 2 and 3 inclusively. It seems very unlikely for a reasonably diverse group of paths to encounter a significantly different congestion distribution in the ``middle mile" than the rest but let us entertain this possibility anyway. 

It is not obvious what signature such a scenario might produce in the resulting analysis, since each ISP could potentially have a different false negative rate determined by congestion encountered by the website paths in its ``middle mile". Assuming that initial segment congestion and ``middle mile" congestion are independent, and given that by hypothesis the former determines the number of recurrently congested connections and the latter the false negative rate for detection of tight initial segments, one would expect there to be little correlation between prevalence of tight initial segments and recurrently congested connections. 

\begin{figure}[h]
\begin{center}
\includegraphics[width=8.5cm]{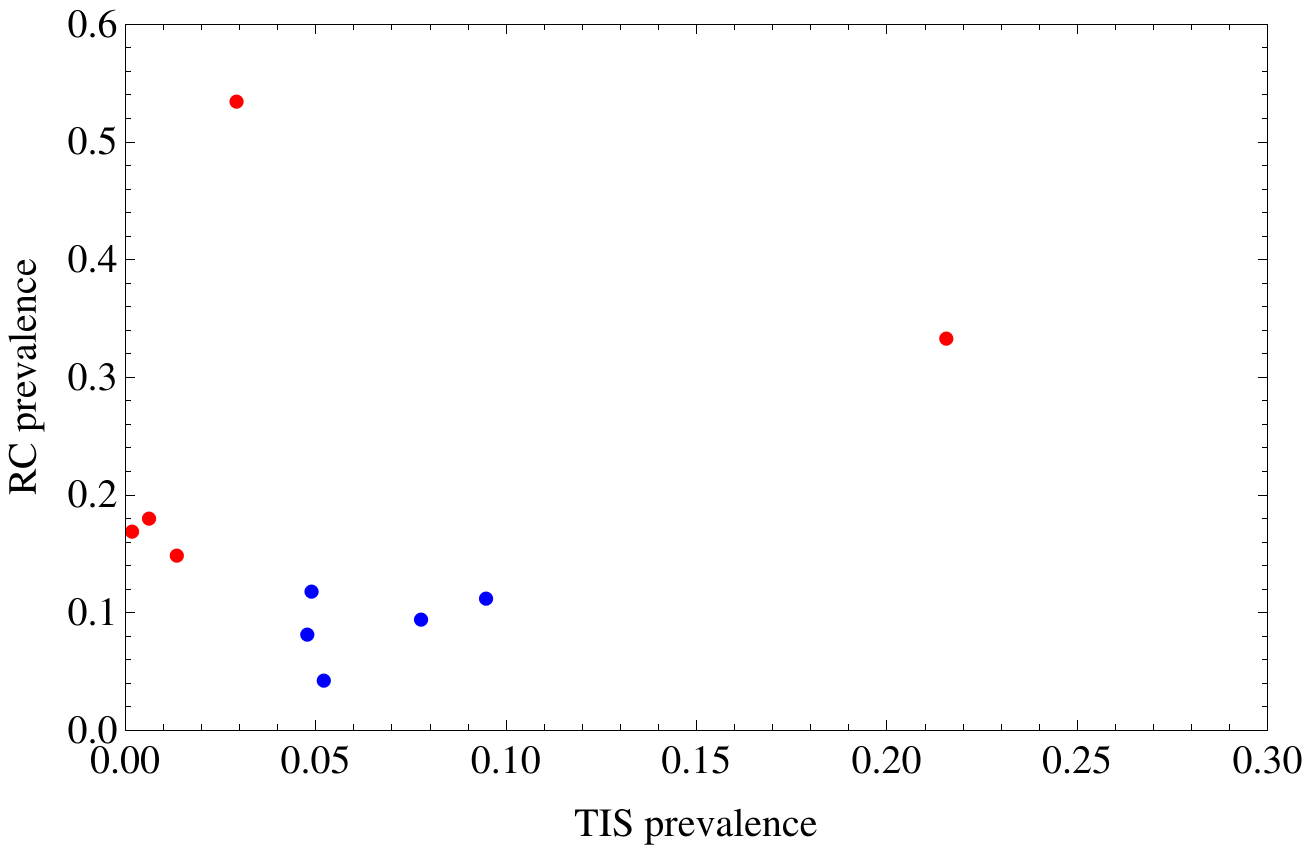}
\end{center}
\caption{Tight initial segment prevalence vs. recurrent congestion prevalence by ISP for the month of April (DSL in blue and cable in red); plot omits outlier ISP 3.}
\label{fig_tis_vs_rc}	
\end{figure}

Figure \ref{fig_tis_vs_rc} shows the plot of recurrent congestion prevalence versus tight initial segment prevalence per ISP for the month of April (other months show a similar distribution of data points). The data has to be considered on the monthly basis because measurements for the same ISP during different months are not independent. The plot exhibits no clear correlation structure. Thus, we cannot rule out that our tight initial segment detection method is foiled by congestion in the ``middle mile".

Figures \ref{fig_dsl_rc_prevalence_by_isp} and \ref{fig_cable_rc_prevalence_by_isp} show plots of recurrent congestion prevalence across ISPs. The plot for DSL connections shows consistently low prevalence of recurrently congested connections. By contrast, the cable plot reveals a great deal of variation between ISPs. Among the ISPs with the largest prevalence (3, 7 and 10), ISPs 3 and 10 had unusually high tight initial segment prevalence, but  the same is not true for 7. This suggests that cable ISPs may have very different distributions of congestion in their networks. Assuming our tight initial segment detection method is accurate, it appears that ISPs 3 and 10 have most of their congestion concentrated on the initial segment while ISP 7 suffers mainly from congestion in the ``middle mile" or deeper.

Based on the above we conclude:
\begin{itemize}
\item Distribution of recurrent congestion differs substantially between DSL and cable broadband access networks.
\item DSL connections experiencing recurrent congestion are constrained in significant numbers on the initial segment but about a third appear to be limited by congestion beyond the initial segment, most likely in the ``middle mile" or beyond.
\item The majority of cable connections experiencing recurrent congestion are constrained beyond the initial segment, most likely in the ``middle mile" or beyond, with a minority, located in networks with particularly high prevalence of recurrent congestion, being limited on the initial segment.
\item Overall, the majority of broadband connections experiencing recurrent congestion appear to be constrained beyond the initial segment.
\end{itemize}
%%%%%%%%%%%%%%%%%%%%%%%%%%%%%%%%%%%%%%%%%%%%%%%%%%%%%%%%%%%%%%%%%%%%%%%%%%%%%%%%%%%%%%%%%%%%%%%%%%%
\section{Open issues}
\label{discussion}

One of the very difficult challenges we hope to address in future publications, is the quantification of uncertainty. While similar approaches have been shown to yield good results in simulation studies \cite{RubKurTow},\cite{HarBesBye},\cite{PadQiuWan},\cite{Duf}, they were typically implemented on time periods of minutes or hours with measurements performed at much higher frequencies than in our study. The longer time scale and the comparatively large number of paths used in our correlation analysis may actually provide a more robust platform for drawing conclusions about congestion since the resulting metrics are less likely to be susceptible to sporadic fluctuations in network performance.  Assuming that the performance of the underlying networks does not change significantly in the course of a few months, the variation in the metrics over the four months considered in the study can be used to quantify measurement uncertainty of the applied methodology \cite{JCG}. Effect of parameter selection on the results can be investigated by performing sensitivity analysis.

An approach similar to the one we presented could also be used to identify recurrently congested segments near the test websites by correlating measurements made for the same website from a large number of different measurement end-points. In the present study we avoided considering correlation between website measurements to avoid potential confusion between correlations due to the user activity cycle and the initial segment congestion. However, by carefully combining the presented data with correlation measurements for websites, we believe it would be possible to reduce the false negative rate for both sets of measurements as well as increase the precision with which congestion can be localized.

Finally, more needs to be done to elucidate the nature of the initial segment of a connection as defined in Section \ref{data_and_definitions} to improve the accuracy of congestion localization. This can potentially be accomplished by analyzing existing \emph{traceroute} data sets such as \cite{GilArlLiMah} and perhaps by less direct methods such as using SamKnows measurement data from secondary measurement servers positioned inside ISP networks.

\section{Conclusion}
\label{conclusion}

Using SamKnows's broadband measurement data, we explored the distribution of congestion in the US Internet broadband access networks. We found big differences in the distribution of systemic congestion between DSL and cable providers. While, DSL ISP networks suffer predominantly from congestion in the ``last mile", distribution of congestion in cable ISP networks exhibits a great deal of variability, with a few cable ISP networks congested mainly in the ``last mile" but the majority congested elsewhere, in the ``middle mile" or beyond.

\bibliographystyle{plain}
\bibliography{samknows_congestion}

\end{document}